\documentclass[pdflatex,sn-mathphys-num]{sn-jnl}

\usepackage{graphicx}%
\usepackage{multirow}%
\usepackage{amsmath,amssymb,amsfonts}%
\usepackage{amsthm}%
\usepackage{mathrsfs}%
\usepackage[title]{appendix}%
\usepackage{xcolor}%
\usepackage{textcomp}%
\usepackage{manyfoot}%
\usepackage{booktabs}%
\usepackage{algorithm}%
\usepackage{algorithmicx}%
\usepackage{algpseudocode}%
\usepackage{listings}%

\def\4He{$^4$He}
\newcommand{\Tl}{\ensuremath{T_\lambda}}
\newcommand{\TKT}{\ensuremath{T_\mathrm{KT}}}
\newcommand{\TKTc}{\ensuremath{T_\mathrm{KT}^\mathrm{corr}}}

\newcommand{\rhosb}{\ensuremath{\rho_\mathrm{sb}}}
\newcommand{\rhosc}{\ensuremath{\rho_\mathrm{sc}}}
\newcommand{\dd}{\ensuremath{\operatorname{d}}}

\begin{document}

\title{Kosterlitz-Thouless transition in uniformly confined $^4$He.}


\author*[1]{\fnm{F.} \sur{Novotný}}\email{filip.novotny@mff.cuni.cz}

\author[1]{\fnm{M.} \sur{Talíř}}

\author[1]{\fnm{B.} \sur{Szalai}}

\author*[1]{\fnm{E.} \sur{Varga}}\email{emil.varga@matfyz.cuni.cz}

\affil*[1]{\orgdiv{Faculty of Mathematics and Physics}, \orgname{Charles University}, \orgaddress{\street{Ke Karlovu 3}, \city{Prague}, \postcode{121 16}, \country{Czech Republic}}}


\abstract{This study investigates the Kosterlitz-Thouless (KT) transition in superfluid $^4$He confined within uniform nanochannels. While the universal jump in superfluid density is a well-established phenomenon, predicting the absolute transition temperature ($T_{KT}$) based on film geometry has remained a long-standing challenge, often relying on empirical fits. Using on-chip nanofluidic Helmholtz resonators with channel heights of 10, 15, and 20 nm, we probe the transition using 4th sound resonant modes.We demonstrate that the observed shift in the transition temperature relative to the bulk lambda point ($T_{\lambda}$) is accurately accounted for by including two-dimensional thermal excitations, specifically 2D rotons. By incorporating these roton-like excitations into the static KT theory, we can predict absolute transition temperatures that align with our experimental measurements and historical data without invoking traditional coherence length scaling arguments. Furthermore, we show that the dynamical extension of the KT theory (AHNS) fully describes the dissipation peaks observed near the transition without requiring ad-hoc free vortex contributions. These results provide compelling evidence that roton excitations, rather than correlation length scaling, govern the finite-size behaviour of confined superfluid $^4$He}


\maketitle

\section{Introduction}\label{sec1}
Most bulk materials cooled to sufficiently low temperatures will transition into an ordered state, e.g., possessing non-zero spontaneous magnetization or macroscopically coherent phase of a macroscopic wave function. Two-dimensional (2D) systems, however, should not have phase transitions at finite temperatures since, according to the Mermin-Wagner theorem \cite{nishimori_elements_2010}, long-wavelength thermal excitations destroy any long-range order. However, it was shown by Kosterlitz and Thouless (KT) \cite{kosterlitz_1973,kosterlitz_1974}, that a type of transition can exist in 2D systems belonging to the XY universality class, such as thin-film superconductors \cite{weitzel_2023}, 2D magnets \cite{hirakawa1982kosterlitz}, or thin layers of \4He \cite{bishop_1978}. In the low-temperature ordered phase, the correlation function $\langle\psi^*(0)\psi(\mathbf{r})\rangle$, with $\psi$ the order parameter, decays with distance slower than exponentially, i.e., as $r^{-\eta(T)}$. This \emph{quasi-long range} order is destroyed at the critical temperature by unbinding of thermally excited bound pairs of vortices and anti-vortices.

Superfluid helium-4 is a nearly ideal material for studying phase transition phenomena in reduced dimensionalities since the superfluid order parameter, the macroscopic wave function, does not couple to the substrate \cite{gasparini_2008}. Its behaviour can be well approximated with a two-fluid model with $\rho = \rho_s + \rho_n$, where $\rho$ is the total density, $\rho_n$ is the viscous normal density and $\rho_s \propto |\psi|^2$ is the superfluid density. In this context, Nelson and Kosterlitz predict \cite{nelson_1977} a transition temperature $\TKT < \Tl \approx 2.1768$~K ($\Tl$ being the bulk transition temperature) and a universal jump in measured superfluid density from $\rho_{s}^{2D}(\TKT)$ to zero as
\begin{equation}
    \label{eq:KT_static}
    \lim_{T\to\TKT}\dfrac{\rho_{s}^{2D}(T)}{T} = 8 \pi k_\mathrm{B} \left(\frac{m}{h}\right)^2,
\end{equation}
where $k_B$ is the Boltzmann constant, $m$ the mass of the \4He atom and $h$ is the Planck constant. The value of this ratio was robustly tested experimentally for very thin films \cite{bishop_1978,bishop_1980,adams_vortex_1987}, although the calculation of the absolute transition temperature from the film geometry is generally not attempted. For relatively thick films (i.e., above 10 nm), on the other hand, the shift in superfluid onset temperature is usually framed in the context of coherence-length scaling of finite size effects \cite{gasparini_2008} with only empirical fits available. In this work, we show that this distinction is unnecessary and that the static Kosterlitz-Thouless theory predicts absolute transition temperature of a superfluid films based only on its thickness if two-dimensional roton-like excitations \cite{varga_2022} are taken into account, which we demonstrate this with measurements and historical data \cite{gasparini_2008}.

The KT transition was widely studied in \4He films using torsional oscillators where the superfluid density of an adsorbed film changes the effective mass \cite{bishop_1978,bishop_1980,agnolet_1989,yano_1999,shirahama1_1990,shirahama2_1990} or using fourth sound (i.e., sound with viscously immobilized normal fluid component \cite{tilley_2003}) acoustic resonators, driven either thermally (adiabatic fountain resonance) \cite{perron_2010,perron_2012,perron_2013} or, more recently, mechanically and fully on-chip \cite{varga_2022}. Experiments probe the superfluid film at finite frequencies, where a subset of the vortices does not relax sufficiently rapidly to orient with the flow. This leads to a broadening of the sharp transition and emergence of a dissipation peak \cite{ambegaokar_1978,ambegaokar_1980,agnolet_1989} (AHNS theory). We measured the peak in acoustic dissipation close to $\TKT$ and we provide the first quantitative test of the AHNS theory near the bulk superfluid $\lambda$ transition.

\section{Results}
\begin{figure}
    \centering
    \includegraphics{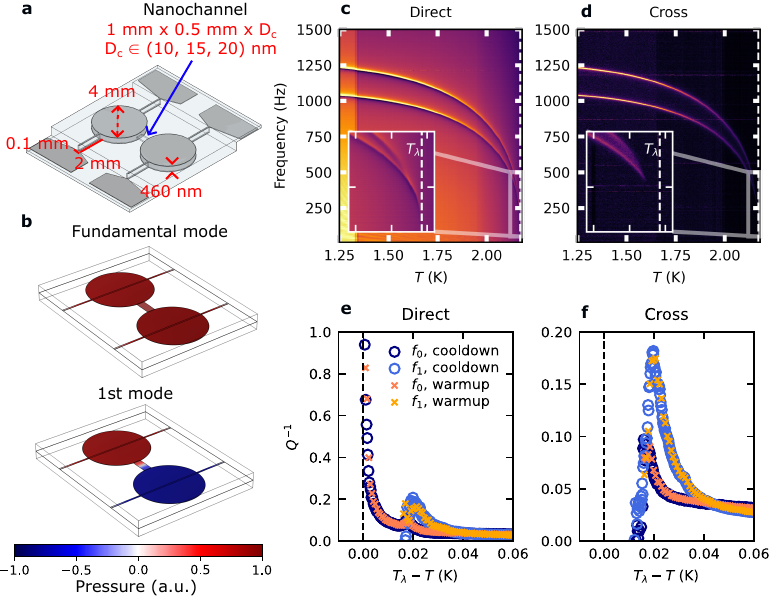}
    \caption{(a) Sketch of the Helmholtz resonator used for the experiment with all relevant dimensions, $D_c$ is the vertical confinement of the nanochannel. (b) Finite elements method numerical simulation of 4th sound pressure for the first two resonance modes in the nanofluidic cavity. (c,d) Heat maps of Direct and Cross signal showing the dependence of the two resonant modes on the temperature, the brightness of the colour marks the amplitude of the signal. Here for $D_c$ = 15~nm. (e,f) Dissipation of both resonant modes for both signals close to the the lambda point for heating up and cooldown characterized by the inverse quality factor. See the presence of the dissipation peaks at $\approx0.02$~K. Here for $D_c$ = 10~nm.}
    \label{fig:meas}
\end{figure}

To study the KT transition we used three nanofluidic Helmholtz resonators with geometries similar to ref.~\cite{rojas_2015,souris_2017,varga_2022}. The resonators consist of two volumes (the circular basin with two filling channels) with height $\approx$460~nm and a connecting nanochannel, with heights 10~nm, 15~nm and 20~nm, see Fig.\ref{fig:meas}(a). The device supports two fourth sound resonant modes of interest (Fig.~\ref{fig:meas}b): a fundamental mode with no flow through the nanochannel and a 1st antisymmetric mode with superflow through the channel. The KT transition in the nanochannel occurs at temperatures sufficiently below the bulk superfluid transition temperature $\Tl$ where it can be well resolved by our thermometry. 

The resonance is excited by electrostatic force acting on one of the basins and the response is monitored on both basins simultaneously, yielding the \emph{direct}, measured on the driven basin, and \emph{cross} response measured on the opposite basin; see details on the experimental arrangement and the detection circuit in the SI. The typical response of the resonator as a function of temperature can be seen in Fig. \ref{fig:meas}(c,d) for both direct and cross signals. One can see that the two modes merge, coincidentally with cross signal disappearing, at certain $T < \Tl$, where the unbinding of vortices destroys the quasi-long range order of superfluid \4He in the nanochannel and the 4th sound can no longer propagate.

The bulk-like and confined superfluid densities are extracted from the resonant response of the devices. The effect of the thermally excited vortices near the phase transition is typically accounted for by introducing a ``dielectric constant'' $\epsilon$ which renormalizes the superfluid density $\rho_s \to \rho_s/\epsilon$. To include this effect we modify the analysis of ref.~\cite{varga_2022}: 

Denoting $\delta p_{1,2}$, $y_{1,2}$, respectively, the pressure fluctuation in the basins and fluid displacements in the filling channels connected to these basins, the equations of motion are \cite{varga_2022}
\begin{align}
    2\rhosb a l \ddot{y}_1 + 2\dfrac{\rhosb}{\rho} a \delta p_1 & = 0, \label{eq:em1} \\
    2\rhosb a l \ddot{y}_2 + 2\dfrac{\rhosb}{\rho} a \delta p_2 & = 0, \label{eq:em2}, \\
\end{align}
where $\rhosb$ is the bulk superfluid density and $a, l$ are the cross section and length of the filling channels. For third sound, AHNS show \cite{ambegaokar_1980} that $\epsilon \dot{\bf{v}}_s = \nabla \mu$ where $\mu$ is the chemical potential. Assuming isothermal conditions in the chip, this translates to the equation of motion for the central channel (displacement $y_c$)
\begin{equation}
    \label{eq:em3}
    \epsilon\rhosc^0 a_c l_c \ddot{y}_c - \dfrac{\rhosc^0}{\rho} (\delta p_1 - \delta p_2)  = 0,
\end{equation}
where $a_c, l_c$ are the cross section and length of the nanochannel. $\rhosc^0$ is the bare superfluid density in the nanochannel unaffected by the quantized vortices. The basin pressures can be then expressed as \cite{varga_2022}
\begin{align}
    \delta p_1 &= k_p \dfrac{2a\rhosb y_1-a_c\rhosc^0y_c}{\rho(2A^2 + V_B\chi k_p)}, \label{eq:p1} \\
    \delta p_2 &= k_p \dfrac{2a\rhosb y_2+a_c\rhosc^0y_c}{\rho(2A^2 + V_B\chi k_p)}. \label{eq:p2}
\end{align}
where $A$ and $V_B$ are the surface area and volume of the basin, $\chi$ is the isothermal compressibility \cite{brooks_1977}. This system of equations of motion has resonance frequencies at (third solution being $\omega_2 = -\omega_1$)
\begin{align}
    \omega_0 &= \sqrt{\dfrac{\rhosb}{\rho^2}\dfrac{2ak_p}{l(2A^2+V_B \chi k_p)}}, \label{eq:f0corr} \\
    \omega_1 &= \sqrt{\dfrac{2ak_p}{l(2A^2 + V_b \chi k_p)}} \sqrt{\dfrac{\rhosb}{\rho^2} + \dfrac{\rhosc^0}{\epsilon\rho^2}\dfrac{a_c l}{a l_c}}. \label{eq:f1corr}
\end{align}

From these the bulk-like and confined superfluid densities are, respectively,
\begin{align}
    \rhosb(T) &= \dfrac{f_0^2(T)}{f_0^2(0)} \rho(T), \label{eq:bulk_den}\\
    \rhosc(T) \equiv \frac{\rhosc^0}{\epsilon} &= \dfrac{f_1^2(T) - f_0^2(T)}{f_1^2(0) - f_0^2(0)} \rho(T) \label{eq:conf_den}
\end{align}
where $f_1(0)$, $f_2(0)$ are frequencies at $T \rightarrow 0$~K, where we assume that $\rhosb = \rhosc^0=\rho$ and $\epsilon \approx1$ and we defined the renormalized confined superfluid density $\rhosc \equiv \rhosc^0/\epsilon$. The zero-temperature limit could not be reached in our experiment (base temperature of our system  $T\approx 1.2$~K).  However, $\rhosc$ is known to  be suppressed with respect to $\rhosb$ by roton-like excitation with energy gap of 5~K \cite{varga_2022,padmore_1974,Kiewiet1975}. We obtain $f_{1,2}(0)$ by fitting the temperature dependence of resonance frequencies to this known scaling (see SI for details). The total density $\rho(T)$ is taken from \cite{donnelly_1998}. 

In addition to the sudden loss of superfluid density, the unbinding of vortex pairs near the KT transition also results in strong dissipation of the superflow, which can be characterized by the inverse quality factor of the resonance mode $Q^{-1} = \gamma/f_{0,1}$. Here the width $\gamma$ and the resonance frequencies $f_{0,1}$ are obtained by the fit to the 4th sound resonances at each temperature. In Fig. \ref{fig:meas}(e,f) we show $Q^{-1}$ against $T$ for the 10~nm resonator for cooling down from $\Tl$ and warming up back towards $\Tl$. The data match well for both type of measurements, suggesting that the system was in an thermal equilibrium during the transition. Further analysis below was performed on the cooldown datasets with minor adjacent averaging to suppress the noise, due to better control over temperature drift on cooldown in our setup.

\begin{figure}
    \centering
    \includegraphics{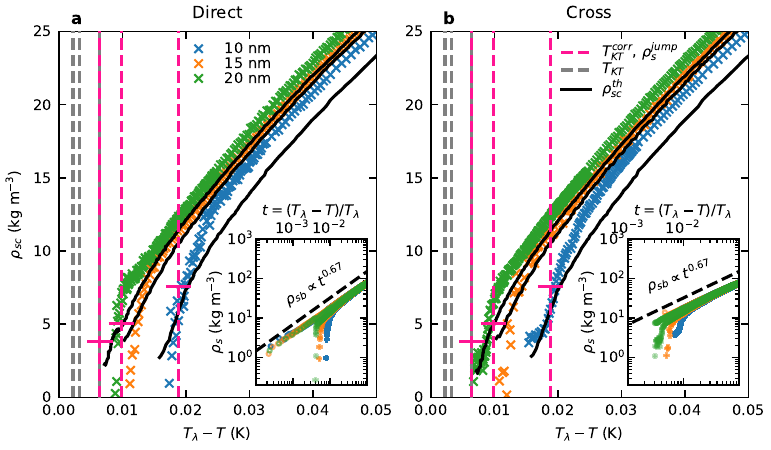}
    \caption{ (a,b) The confined \rhosc~(crosses) superfluid density, obtained using the Eq. \eqref{eq:conf_den} against the shifted temperature $T_{\lambda} - T$ for both signals and all three resonators. The grey lines mark the $\TKT$ calculated from the Eq.\eqref{eq:KT_static}, from the left (20~nm) to the right (10~nm), the vertical pink dashed lines then correspond to the $\TKTc$ and horizontal to the universal jump in superfluid density at $\TKTc$. The black lines show the superfluid density renormalized by the dielectric constant $\epsilon$ obtained from the fit of dissipation. The insets then show both the $\rhosb$ and $\rhosc$ against reduced temperature $t = (T_{\lambda} - T)/T_{\lambda}$, the bulk density follows well the tabulated critical behaviour $\rhosb \propto t^{-0.67}$.}
    \label{fig:rhos_vs_t}
\end{figure}

The confined superfluid density calculated using the Eq.\eqref{eq:conf_den} is plotted against the shifted temperature $T_{\lambda} - T$ in Fig.\ref{fig:rhos_vs_t}(a,b) for all three confinements. The bulk data, shown in the insets overlap each other and follow well the high temperature behaviour of bulk superfluid density $\rho_s \propto |t|^{0.67}$ \cite{singasaas_1984}, with $t=(\Tl - T)/\Tl$ the reduced temperature. Uncertainty in absolute temperature is corrected by a fit of the ratio $\rhosb/\rho$ against temperature to known behaviour of bulk superfluid density to obtain $\Tl$ (see the SI). The confined density, on the other hand, shows a sudden and steep drop at $T < \Tl$, which is characteristic of the KT transition \cite{bishop_1980}. The cross measurements are cut above certain temperatures, since at $T \simeq \TKT$ the cross signal disappears completely due to the transition of the nanochannel. 

The two dimensional superfluid density entering \eqref{eq:KT_static} is the renormalized $\rho_s^\mathrm{2D} = \rho_s^{\mathrm{2D}0}/\epsilon$, where $\rho_s^{\mathrm{2D}0}$ is the bare superfluid 2D density (i.e., without vortices) obtained by the integration of the bare 3D density $\rhosb$ over the confined dimension, i.e., $\rho_{s}^{2D0} = \int \rhosb \dd z \approx \rhosb D_c$. The temperature-dependent $\epsilon \approx 1$ for temperatures sufficiently below $\TKT$ and, in the limit of zero velocity and frequency, $\epsilon = \infty$ for $T > \TKT$. In the static case, we neglect the effect of vortices away from the narrow region of the transition, and solve \eqref{eq:KT_static} numerically for a given $D_c$ with an approximation $\rho_{s}^{2D} \approx \rhosb D_c$, where for the bulk superfluid density the $\rhosb$ measured with the fundamental mode was used. This yields $\TKT$ indicated by the gray dashed vertical lines in Fig.~\ref{fig:rhos_vs_t}, which significantly over-estimate the observed transition temperatures.

In addition to the effect of the quantized vortices, the superfluid density in the nanochannel is also suppressed compared to the bulk superfluid density by the 2D roton excitations \cite{padmore_1974,varga_2022}, which contribute to the 2D normal density as
\begin{equation}\label{eq:rho_roton}
    \Delta\rho_\mathrm{n}^{2D} = \dfrac{\hbar k_0^3 m^{*1/2}}{2 \sqrt{2 \pi k_B T}} \exp{\left( -\dfrac{\Delta}{k_B T} \right)},
\end{equation}
where $\Delta/k_B \approx 5$~K is the 2D roton gap, $k_0 \approx 1.8~\textrm{\AA}^{-1}$ the roton minimum wave vector and $m^{*} = 0.20 m_{4\textrm{He}}$ the roton effective mass \cite{padmore_1974,chester_1976,arrigoni_2013}.

\begin{figure}
    \centering
    \includegraphics{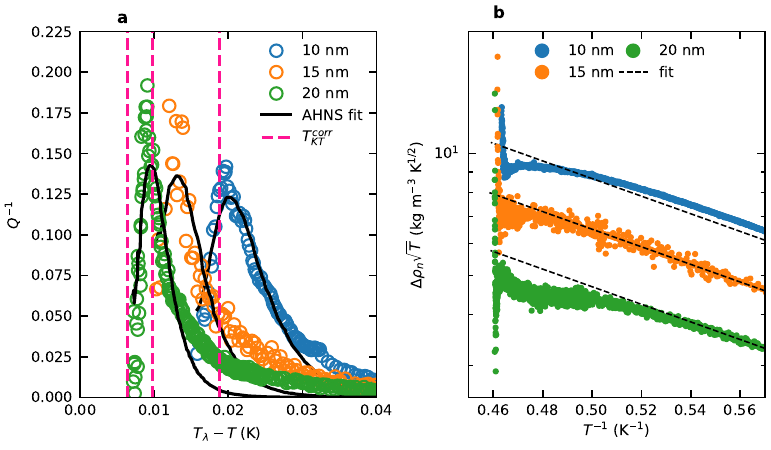}
    \caption{(a) The inverse quality factor $Q^{-1}$ against the shifter temperature $T_{\lambda} - T$ for all three film thickness. $Q^{-1}$ characterizes the dissipation of the resonance mode during the KT-transition. The black lines are fits of the AHNS theory and the pink dashed lines correspond to the $\TKTc$. (b) The measured \rhosc against $T_{\lambda} - T$ for the cross signal. The dashed lines are calculated confined densities using the tabulated bulk superfluid density and real part of the dielectric constant obtained from the fits in Fig. (a).} 
    \label{fig:diss}
\end{figure}

In the same approximation as before, i.e., neglecting the contribution of vortices outside of a region very close to $\TKT$, the superfluid density entering \eqref{eq:KT_static} is $\rho_s^{2D} = \rhosc^0 D_c = \rhosb D_c - \Delta\rho_\mathrm{n}^{2D}$, where $\rhosc^0$ is the average 3D \emph{bare} superfluid density in the nanochannel. Solving \eqref{eq:KT_static} numerically with this $\rho_s^{2D}$ yields the corrected transition temperatures $\TKTc$. Corrected transition temperatures $\TKTc$ together with the universal jumps in $\rhosc$ are marked by the vertical (horizontal) pink dashed lines in Fig.~\ref{fig:rhos_vs_t} and Fig.~\ref{fig:diss}a and are tabulated in Tab.~\ref{tab:Tc}. The calculated transition temperatures are in good agreement with our measurement. With the roton contribution taken into account, the corrected transition temperatures $\TKTc$, lie close the region of rapid $\rhosc$ drop (Fig.~\ref{fig:rhos_vs_t}) and the peak in dissipation (Fig.~\ref{fig:diss}a).

In the more realistic case of finite velocities and frequencies the theory must be extended to include the dynamics of vortices and their finite response time. AHNS showed \cite{ambegaokar_1978,ambegaokar_1980,agnolet_1989} that in the dynamical case, the dielectric constant takes complex values. Experimentally, this manifests as peak in the dissipation, usually characterized by the inverse quality factor $Q^{-1}$ of the used resonator, i.e., torsional oscillator with \4He film on an oscillating substrate \cite{bishop_1978,bishop_1980,agnolet_1989,yano_1999}, or, in our case, the antisymmetric Helmholtz mode, whose quality factor is estimated from \eqref{eq:f1corr} as 
\begin{equation}\label{eq:invQdef}
    Q_{KT}^{-1} = \left( \dfrac{\textrm{Re}\{ \omega_1\}}{\textrm{Im}\{ \omega_1 \}} \right)^{-1}.
\end{equation}
Assuming that $(\rhosc^0 a_c l)^2 \ll (\rhosb a l_c)^2$ (since $a_c \ll a$) and keeping only leading order of $\epsilon^{-1}$ (since $|\epsilon|\to\infty$ as the transition temperature is approached), \eqref{eq:invQdef} can be simplified to
\begin{equation} \label{eq:invQ_KT}
    Q_{KT}^{-1} \approx \dfrac{\textrm{Im}\{ \epsilon\}}{|\epsilon|^2} \dfrac{\rhosc^0 a_c l}{\rhosb a l_c},
\end{equation}
which is equivalent to the result derived for oscillating substrates $Q^{-1} \propto \mathrm{Im}[\epsilon^{-1}]$ \cite{bishop_1978,bishop_1980,ambegaokar_1980}, albeit with a different prefactor, which is of order $a_c/a \approx 10^{-1}$ (in torsional oscillators the prefactor $\rho_\mathrm{s}^{2D} A / M \approx 10^{-6}$ \cite{bishop_1980}, where $A$ is the area of the substrate, $M$ the effective mass of the oscillator and $\rho_\mathrm{s}^{2D}$ the renormalized 2D superfluid density).

The experimentally observed peak width, however, includes the contributions from the nanochannel itself and residual viscous flow in the four inlets. The $f_0$ mode, on the other hand, dissipates primarily via the viscous flow in the inlets (see Fig.~\ref{fig:meas}b). Thus, to isolate the dissipation due to the nanochannel we have
\begin{equation}\label{eq:invQ_exp}
    Q^{-1} = Q_{1c}^{-1} - Q_{0d}^{-1}\dfrac{f_{0d}}{f_{1c}},
\end{equation}
where $f_{0d}$ is the frequency of the $f_0$ mode and $f_{1c}$ of the $f_1$ mode and subscripts c and d refer to direct and cross measurements, respectively. The rescaling takes into account the fact that the dissipation of the modes (i.e., resonance width) is equal for both modes at sufficiently low $T$ since it stems from the residual viscous flow in the inlet channels, but the frequencies are not (see SI for additional details). Note that $f_{0d}$ mode also displays a small dissipation peak near $\TKT$ due to asymmetry of the forcing. This small peak was cut and interpolated over by a power law fit, causing a slight perturbation of the high-$T$ side of the peak in $Q^{-1}$. Temperature dependence of the resulting $Q^{-1}$ calculated using Eq.\eqref{eq:invQ_exp} is shown in Fig.~\ref{fig:diss}(a) together with $\TKTc$. 

The fit of \eqref{eq:invQ_KT} to the measured data is shown with full black lines in Fig.~\ref{fig:diss}(a). The value of $\epsilon $ is given by the numerical integration of the KT recursion relations \cite{kosterlitz_1973,kosterlitz_1974,ambegaokar_1980} (see Methods). The principal quantity that determines the finite-frequency response is the vortex diffusivity $D$ which determines the maximum scale of the recursion relations as $l_\mathrm{max} = \log(\sqrt{14 D} / \sqrt\omega a_0)$, i.e., the recursion relations are truncated at vortex pair separation larger than $a_0e^{l_\mathrm{max}}$. We determine the transition temperature $\TKT^\mathrm{fit}$ from $K(l_\mathrm{max}) = 2/\pi$, where $K$ is the scale-dependent superfluid stiffness (see Methods). These are shown in Tab.~\ref{tab:Tc} and are in good agreement with the static estimate $\TKTc$. The diffusivity $D$ obtained from the fit is also listed in Tab.~\ref{tab:Tc}. We find the diffusivity to be significantly lower than previously obtained results, which are of the order of $10^{-4}$~cm$^2$s$^{-1}$, but continue the decreasing trend with increasing temperature \cite{bishop_1980,agnolet_thermal_1981,adams_vortex_1987}, although we note that $D$ is expected to blow up \cite{adams_vortex_1987} near $\TKT$ and thus the constant $D$ used here is only an approximation. This is likely responsible for the imperfect fit of the dissipation peak tails seen in Fig.~\ref{fig:diss}a. We also note that the $D$ was calculated assuming vortex core parameter $a_0\approx 1.5\times 10^{-8}$~cm, if the coherence length $\xi\approx 3.5$~nm at $T\approx 2.16$~K is used in place of $a_0$, the diffusivities increase by about 2 orders of magnitude. This dramatic reduction in $D$ is most likely a consequence of the extreme vertical confinement: atomic force microscopy (AFM) of our nanochannels reveals an RMS surface roughness of approximately 1 to 2 nm (see the SI). In a 10 nm to 20 nm channel, this roughness constitutes up to 20\% of the total fluid height, creating a strong geometric pinning potential that immobilizes vortices, a phenomenon qualitatively in line with recent observations of turbulence decay in confined geometries \cite{novotny_2026}. Notably, however, we find that the dissipation peak can be satisfactorily fit with the bound-vortex $\epsilon_b$ \cite{ambegaokar_1980}, which arises from the truncation of the KT recursion relations, without the addition of the free vortex contribution $\epsilon_f$ required to explain the dissipation peak in torsional pendulum experiments \cite{ambegaokar_1980,bishop_1978,bishop_1980}.

The dielectric constant $\epsilon$ obtained from the fit to the dissipation can be used to calculate the renormalized confined superfluid density, i.e., $\rhosc = \rhosc^0/\mathrm{Re}\{\epsilon\}$, where $\rhosc^0$ is the bare density used for calculation of the static $\TKT$. The resulting temperature dependence of superfluid densities is shown in Fig.~\ref{fig:rhos_vs_t} as black lines. These slightly underestimate the measured confined superfluid density $\rhosc$. The origin of this under-estimation can be seen in Fig.~\ref{fig:diss}b, which shows the excess normal fluid density $\rhosc - \rhosb$ (where the densities are the measured densities given by \eqref{eq:bulk_den} and \eqref{eq:conf_den}). The extrapolation of the 2D roton contribution to the normal fluid density overestimates the excess normal fluid density above approximately 2 K for all three confinements. Finite-size effects generally result in further suppression of superfluid density \cite{gasparini_2008}. The \emph{enhancement} of superfluid density in confined geometries, however, is in qualitative agreement with vortex-loop model of superfluid transition \cite{Williams1987}. In this case, the 3D superfluid transition is driven by thermal excitation of vortex loops, as opposed to vortex dipoles in the 2D case. In finite geometries, the size of the largest loops is truncated by the confinement, thus resulting in increased superfluid stiffness \cite{Williams1987}.

\begin{table}
    \centering
    \begin{tabular}{ccccc}
    \hline
        $D_c$ (nm) & $T_{\lambda}-\TKT$ (mK) & $T_{\lambda}-\TKTc$ (mK) & $T_{\lambda}-T^{fit}_{KT}$ (mK) & $D$ (cm$^2$s$^{-1}$) \\
        \hline
        10 & 6.4 & 18.8 & 20.1 & 7.6 $\times$ 10$^{-12}$\\
        15 & 3.2 & 9.8 & 11.5 & 9.0 $\times$ 10$^{-12}$\\
        20 & 2.2 & 6.4 & 8.7 & 3.2 $\times$ 10$^{-12}$\\
    \end{tabular}
    \caption{The shift of superfluid transition temperature in the nanochannel for all three confinements with respect to $\Tl$. $\TKT$ is calculated from the \eqref{eq:KT_static}, $\TKTc$ includes the superfluid density suppression due to the roton excitations and $T^{fit}_{KT}$ is obtained from the AHNS theory fit to the data. In the last column we show the diffusivity constants obtained by the AHNS theory fit assuming that the vortex core parameter $a_0 = 1.5\times 10^{-8}$~cm.}
    \label{tab:Tc}
\end{table}

\begin{figure}
    \centering
    \includegraphics{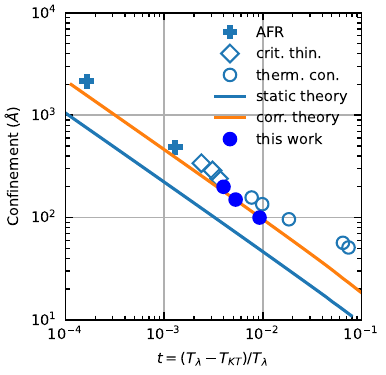}
    \caption{Superfluid phase transition temperatures as a function of confinement. The static $\TKT$ \eqref{eq:KT_static} (blue line) corrected for superfluid density suppression due to roton excitations (orange line) accounts for nearly all observed shift of the transition temperature with slab thickness, with no adjustable parameters. Historical data taken from Fig.~24 of the review \cite{gasparini_2008}, which contains data from refs.~\cite{gasparini_1998,kimball_2001,yu_1989,ganshin_2006}.}
    \label{fig:gasparini-data}
\end{figure}

The shift of the superfluid onset temperature with film thickness, for relatively thick films (compared to near-monolayers of, e.g., \cite{bishop_1980}), is typically framed as a finite size effect that scales with the coherence length \cite{gasparini_2008}. Our data indicate that the shift of the superfluid transition temperature with film thickness can be nearly completely accounted for by static Nelson-Kosterlitz result \eqref{eq:KT_static} corrected for 2D roton excitations. To further substantiate this claim we compare the calculated $\TKTc$ with previously reported critical temperatures obtained under various confinements using adiabatic fountain resonance \cite{gasparini_1998,kimball_2001}, thermal conductivity \cite{yu_1989}, and critical thinning measurements \cite{ganshin_2006} as shown in Fig.\ref{fig:gasparini-data}. We find good agreement with historical data without invoking coherence length scaling arguments \cite{gasparini_2008} and with no adjustable parameters.

\section{Conclusions}
Using in-situ comparison of bulk and confined superfluid densities in $^4$He close to the superfluid transition temperature, we show that the static Kosterlitz-Thouless theory can predict absolute transition temperatures if enhancement of the normal fluid density due to roton-like excitations is taken into account. We show that this correction fully accounts for the observed transition temperatures in thin films in present and past experiments, which relied on a variety of experimental techniques. We also show that the dynamical extension of the Kosterlitz-Thouless theory fully accounts for the dissipation peak in the confined channel without the ad-hoc addition of the free vortex contribution. The presented work constitutes further evidence that the correlation length scaling usually invoked to explain finite-size scaling in superfluid helium \cite{gasparini_2008} cannot account for the behaviour of confined superfluid $^4$He.

\section{Methods}

To calculate the finite-frequency response of the superfluid, including corrections for finite velocities, we numerically integrate the Kosterlitz-Thouless recursion relations, mostly following ref.~\cite{adams_vortex_1987}
\begin{equation}\label{K1}
    \dfrac{\dd K}{\dd \ell} = -4\pi^3 K^2(\ell) y^2(\ell),
\end{equation}
\begin{equation}\label{K2}
    \dfrac{\dd y}{\dd \ell} = \left( 2 - \pi K(\ell) + \frac{1}{2} z(\ell) \frac{I_1[z(\ell)]}{I_0[z(\ell)]}\right)y(\ell),
\end{equation}
from $\ell = 0$ and where $K(\ell)$, $y(\ell)$ are the scale-dependent superfluid stiffness and vortex fugacity. The initial conditions $K_0 = K(\ell = 0) = (\hbar/m)^2 (\rhosc^{\mathrm{2D}0} / k_B T)$ is the microscopic superfluid stiffness given by the bare superfluid density and $y_0 = e^{-E_c/k_B T}$, where $E_c$ is the energy of the vortex core. The third term in the parenthesis in \eqref{K2} is a finite velocity correction due to superflow of characteristic velocity $V_s$ \cite{adams_vortex_1987}, $I_k$ is the modified Bessel function of order $k$, $z(\ell) = 2\pi K_0 e^\ell V_s/V_0$ and $V_0 = h/(ma_0)$ is of the order of 10~m/s (depending on $a_0$) \cite{adams_vortex_1987}. In our case we used the ratio $V_s/V_0$ as a fitting parameter. 

In the static case, the dielectric constant is given by $\epsilon = K_0/K(\ell\to\infty)$. For the dynamical case, we follow AHNS \cite{ambegaokar_1980} for bound vortices and finite frequencies. The integration runs from from $\ell = 0$ to a certain $\ell_{max} = \ln{(r_D/a_0)}$, which is the logarithm of the ratio between the diffusion length $r_D = \sqrt{14D/\omega}$ and the vortex core radius $a_0 \approx 1.5$~$\textrm{\AA}$ \cite{tilley_2003}. Note, however, that the theory depends only on the ratio $D/a_0^2$.

The real part of the dielectric constant, determining the superfluid density, is
\begin{equation}
    \textrm{Re}\{ \epsilon_b \} = \dfrac{K_0}{K(\ell_\mathrm{max})},
\end{equation}
and the imaginary component, determining the dissipation peak, is
\begin{equation}
    \textrm{Im}\{ \epsilon_b \} = \pi^4 K_0 y^2(\ell_\mathrm{max}),
\end{equation}
where $y(\ell)$ is the scale-dependent vortex fugacity. $K(\ell)$ and $y(\ell)$ are found as the solutions of the Kosterlitz recursion relations.

To perform the fit in Fig.~\ref{fig:diss}, the recursion relations were integrated at each temperature, which enters as a parameter via $\rhosc^{\mathrm{2D}0}$, which is the same as the superfluid density used for calculating the static $\TKTc$. The parameters adjusted by the fit were $D$, $y_0$, $V_s/V_0$ and an overall scaling constant in the expression for $Q^{-1}$ \eqref{eq:invQ_KT}. The frequency was fixed at $\omega = 2\pi \times 300$~Hz, since it changes only negligibly. Including the free-vortex contribution \cite{ambegaokar_1980} was found to produce negligible differences and was not included in the final fit.

\backmatter

\bmhead{Acknowledgements}
The work was supported by Charles University under PRIMUS/23/SCI/017. F. N., in addition, acknowledges the financial support from Charles University under GAUK 129724. CzechNanoLab projects LM2023051 and LNSM-LNSpin funded by MEYS CR are gratefully acknowledged for the financial support of the sample fabrication at CEITEC Nano Research Infrastructure and LNSM at FZU AV\v{C}R.

\section*{Declarations}
The authors have nothing to declare.

\section*{Author Contributions}
F.N. fabricated the resonators and collected the data. F.N. and E.V. wrote the manuscript. M.T. and B.S. developed measurement circuitry and software. E.V. supervised the project.

\section*{Data availability.}
All source data for figures in the main text and supplementary information are available in a Zenodo repository \cite{zenodo}.


\bibliography{sn-bibliography}


\begin{thebibliography}{46}
\ifx \bisbn   \undefined \def \bisbn  #1{ISBN #1}\fi
\ifx \binits  \undefined \def \binits#1{#1}\fi
\ifx \bauthor  \undefined \def \bauthor#1{#1}\fi
\ifx \batitle  \undefined \def \batitle#1{#1}\fi
\ifx \bjtitle  \undefined \def \bjtitle#1{#1}\fi
\ifx \bvolume  \undefined \def \bvolume#1{\textbf{#1}}\fi
\ifx \byear  \undefined \def \byear#1{#1}\fi
\ifx \bissue  \undefined \def \bissue#1{#1}\fi
\ifx \bfpage  \undefined \def \bfpage#1{#1}\fi
\ifx \blpage  \undefined \def \blpage #1{#1}\fi
\ifx \burl  \undefined \def \burl#1{\textsf{#1}}\fi
\ifx \doiurl  \undefined \def \doiurl#1{\url{https://doi.org/#1}}\fi
\ifx \betal  \undefined \def \betal{\textit{et al.}}\fi
\ifx \binstitute  \undefined \def \binstitute#1{#1}\fi
\ifx \binstitutionaled  \undefined \def \binstitutionaled#1{#1}\fi
\ifx \bctitle  \undefined \def \bctitle#1{#1}\fi
\ifx \beditor  \undefined \def \beditor#1{#1}\fi
\ifx \bpublisher  \undefined \def \bpublisher#1{#1}\fi
\ifx \bbtitle  \undefined \def \bbtitle#1{#1}\fi
\ifx \bedition  \undefined \def \bedition#1{#1}\fi
\ifx \bseriesno  \undefined \def \bseriesno#1{#1}\fi
\ifx \blocation  \undefined \def \blocation#1{#1}\fi
\ifx \bsertitle  \undefined \def \bsertitle#1{#1}\fi
\ifx \bsnm \undefined \def \bsnm#1{#1}\fi
\ifx \bsuffix \undefined \def \bsuffix#1{#1}\fi
\ifx \bparticle \undefined \def \bparticle#1{#1}\fi
\ifx \barticle \undefined \def \barticle#1{#1}\fi
\bibcommenthead
\ifx \bconfdate \undefined \def \bconfdate #1{#1}\fi
\ifx \botherref \undefined \def \botherref #1{#1}\fi
\ifx \url \undefined \def \url#1{\textsf{#1}}\fi
\ifx \bchapter \undefined \def \bchapter#1{#1}\fi
\ifx \bbook \undefined \def \bbook#1{#1}\fi
\ifx \bcomment \undefined \def \bcomment#1{#1}\fi
\ifx \oauthor \undefined \def \oauthor#1{#1}\fi
\ifx \citeauthoryear \undefined \def \citeauthoryear#1{#1}\fi
\ifx \endbibitem  \undefined \def \endbibitem {}\fi
\ifx \bconflocation  \undefined \def \bconflocation#1{#1}\fi
\ifx \arxivurl  \undefined \def \arxivurl#1{\textsf{#1}}\fi
\csname PreBibitemsHook\endcsname

\bibitem[\protect\citeauthoryear{Nishimori and Ortiz}{2010}]{nishimori_elements_2010}
\begin{bbook}
\bauthor{\bsnm{Nishimori}, \binits{H.}},
\bauthor{\bsnm{Ortiz}, \binits{G.}}:
\bbtitle{Elements of {Phase} {Transitions} and {Critical} {Phenomena}}.
\bsertitle{Oxford {Graduate} {Texts}}.
\bpublisher{Oxford University Press},
\blocation{Oxford}
(\byear{2010}).
\doiurl{10.1093/acprof:oso/9780199577224.001.0001}
\end{bbook}
\endbibitem

\bibitem[\protect\citeauthoryear{Kosterlitz and Thouless}{1973}]{kosterlitz_1973}
\begin{barticle}
\bauthor{\bsnm{Kosterlitz}, \binits{J.M.}},
\bauthor{\bsnm{Thouless}, \binits{D.J.}}:
\batitle{Ordering, metastability and phase transitions in two-dimensional systems}.
\bjtitle{Journal of Physics C: Solid State Physics}
\bvolume{6}(\bissue{7}),
\bfpage{1181}--\blpage{1203}
(\byear{1973})
\doiurl{10.1088/0022-3719/6/7/010}
\end{barticle}
\endbibitem

\bibitem[\protect\citeauthoryear{Kosterlitz}{1974}]{kosterlitz_1974}
\begin{barticle}
\bauthor{\bsnm{Kosterlitz}, \binits{J.M.}}:
\batitle{The critical properties of the two-dimensional {XY} model}.
\bjtitle{Journal of Physics C: Solid State Physics}
\bvolume{7}(\bissue{6}),
\bfpage{1046}--\blpage{1060}
(\byear{1974})
\doiurl{10.1088/0022-3719/7/6/005}
\end{barticle}
\endbibitem

\bibitem[\protect\citeauthoryear{Weitzel et~al.}{2023}]{weitzel_2023}
\begin{barticle}
\bauthor{\bsnm{Weitzel}, \binits{A.}},
\bauthor{\bsnm{Pfaffinger}, \binits{L.}},
\bauthor{\bsnm{Maccari}, \binits{I.}},
\bauthor{\bsnm{Kronfeldner}, \binits{K.}},
\bauthor{\bsnm{Huber}, \binits{T.}},
\bauthor{\bsnm{Fuchs}, \binits{L.}},
\bauthor{\bsnm{Mallord}, \binits{J.}},
\bauthor{\bsnm{Linzen}, \binits{S.}},
\bauthor{\bsnm{Il’ichev}, \binits{E.}},
\bauthor{\bsnm{Paradiso}, \binits{N.}},
\bauthor{\bsnm{Strunk}, \binits{C.}}:
\batitle{Sharpness of the {Berezinskii}-{Kosterlitz}-{Thouless} {Transition} in {Disordered} {NbN} {Films}}.
\bjtitle{Physical Review Letters}
\bvolume{131}(\bissue{18}),
\bfpage{186002}
(\byear{2023})
\doiurl{10.1103/PhysRevLett.131.186002}
\end{barticle}
\endbibitem

\bibitem[\protect\citeauthoryear{Hirakawa}{1982}]{hirakawa1982kosterlitz}
\begin{barticle}
\bauthor{\bsnm{Hirakawa}, \binits{K.}}:
\batitle{Kosterlitz-thouless transition in two-dimensional planar ferromagnet {K$_2$CuF$_4$}}.
\bjtitle{Journal of Applied Physics}
\bvolume{53}(\bissue{3}),
\bfpage{1893}--\blpage{1898}
(\byear{1982})
\end{barticle}
\endbibitem

\bibitem[\protect\citeauthoryear{Bishop and Reppy}{1978}]{bishop_1978}
\begin{barticle}
\bauthor{\bsnm{Bishop}, \binits{D.J.}},
\bauthor{\bsnm{Reppy}, \binits{J.D.}}:
\batitle{Study of the {Superfluid} {Transition} in {Two}-{Dimensional} {$^4$He} {Films}}.
\bjtitle{Physical Review Letters}
\bvolume{40}(\bissue{26}),
\bfpage{1727}--\blpage{1730}
(\byear{1978})
\doiurl{10.1103/PhysRevLett.40.1727}
\end{barticle}
\endbibitem

\bibitem[\protect\citeauthoryear{Gasparini et~al.}{2008}]{gasparini_2008}
\begin{barticle}
\bauthor{\bsnm{Gasparini}, \binits{F.M.}},
\bauthor{\bsnm{Kimball}, \binits{M.O.}},
\bauthor{\bsnm{Mooney}, \binits{K.P.}},
\bauthor{\bsnm{Diaz-Avila}, \binits{M.}}:
\batitle{Finite-size scaling of {$^4$He} at the superfluid transition}.
\bjtitle{Reviews of Modern Physics}
\bvolume{80}(\bissue{3}),
\bfpage{1009}--\blpage{1059}
(\byear{2008})
\doiurl{10.1103/RevModPhys.80.1009}
\end{barticle}
\endbibitem

\bibitem[\protect\citeauthoryear{Nelson and Kosterlitz}{1977}]{nelson_1977}
\begin{barticle}
\bauthor{\bsnm{Nelson}, \binits{D.R.}},
\bauthor{\bsnm{Kosterlitz}, \binits{J.M.}}:
\batitle{Universal {Jump} in the {Superfluid} {Density} of {Two}-{Dimensional} {Superfluids}}.
\bjtitle{Physical Review Letters}
\bvolume{39}(\bissue{19}),
\bfpage{1201}--\blpage{1205}
(\byear{1977})
\doiurl{10.1103/PhysRevLett.39.1201}
\end{barticle}
\endbibitem

\bibitem[\protect\citeauthoryear{Bishop and Reppy}{1980}]{bishop_1980}
\begin{barticle}
\bauthor{\bsnm{Bishop}, \binits{D.J.}},
\bauthor{\bsnm{Reppy}, \binits{J.D.}}:
\batitle{Study of the superfluid transition in two-dimensional {$^4$He} films}.
\bjtitle{Physical Review B}
\bvolume{22}(\bissue{11}),
\bfpage{5171}--\blpage{5185}
(\byear{1980})
\doiurl{10.1103/PhysRevB.22.5171}
\end{barticle}
\endbibitem

\bibitem[\protect\citeauthoryear{Adams and Glaberson}{1987}]{adams_vortex_1987}
\begin{barticle}
\bauthor{\bsnm{Adams}, \binits{P.W.}},
\bauthor{\bsnm{Glaberson}, \binits{W.I.}}:
\batitle{Vortex dynamics in superfluid helium films}.
\bjtitle{Physical Review B}
\bvolume{35}(\bissue{10}),
\bfpage{4633}--\blpage{4652}
(\byear{1987})
\doiurl{10.1103/PhysRevB.35.4633}
\end{barticle}
\endbibitem

\bibitem[\protect\citeauthoryear{Varga et~al.}{2022}]{varga_2022}
\begin{barticle}
\bauthor{\bsnm{Varga}, \binits{E.}},
\bauthor{\bsnm{Undershute}, \binits{C.}},
\bauthor{\bsnm{Davis}, \binits{J.P.}}:
\batitle{Surface-{Dominated} {Finite}-{Size} {Effects} in {Nanoconfined} {Superfluid} {Helium}}.
\bjtitle{Physical Review Letters}
\bvolume{129}(\bissue{14}),
\bfpage{145301}
(\byear{2022})
\doiurl{10.1103/PhysRevLett.129.145301}
\end{barticle}
\endbibitem

\bibitem[\protect\citeauthoryear{Agnolet et~al.}{1989}]{agnolet_1989}
\begin{barticle}
\bauthor{\bsnm{Agnolet}, \binits{G.}},
\bauthor{\bsnm{McQueeney}, \binits{D.F.}},
\bauthor{\bsnm{Reppy}, \binits{J.D.}}:
\batitle{Kosterlitz-{Thouless} transition in helium films}.
\bjtitle{Physical Review B}
\bvolume{39}(\bissue{13}),
\bfpage{8934}--\blpage{8958}
(\byear{1989})
\doiurl{10.1103/PhysRevB.39.8934}
\end{barticle}
\endbibitem

\bibitem[\protect\citeauthoryear{Yano et~al.}{1999}]{yano_1999}
\begin{barticle}
\bauthor{\bsnm{Yano}, \binits{H.}},
\bauthor{\bsnm{Jocha}, \binits{T.}},
\bauthor{\bsnm{Wada}, \binits{N.}}:
\batitle{Vortex-dynamics study of the frequency dependence of the superfluid onset temperature}.
\bjtitle{Physical Review B}
\bvolume{60}(\bissue{1}),
\bfpage{543}--\blpage{549}
(\byear{1999})
\doiurl{10.1103/PhysRevB.60.543}
\end{barticle}
\endbibitem

\bibitem[\protect\citeauthoryear{Shirahama et~al.}{1990a}]{shirahama1_1990}
\begin{barticle}
\bauthor{\bsnm{Shirahama}, \binits{K.}},
\bauthor{\bsnm{Kubota}, \binits{M.}},
\bauthor{\bsnm{Ogawa}, \binits{S.}},
\bauthor{\bsnm{Wada}, \binits{N.}},
\bauthor{\bsnm{Watanabe}, \binits{T.}}:
\batitle{Kosterlitz-{Thouless} superfluid transition in {$^4$He} films adsorbed on porous glasses}.
\bjtitle{Physica B: Condensed Matter}
\bvolume{165-166},
\bfpage{545}--\blpage{546}
(\byear{1990})
\doiurl{10.1016/S0921-4526(90)81122-5}
\end{barticle}
\endbibitem

\bibitem[\protect\citeauthoryear{Shirahama et~al.}{1990b}]{shirahama2_1990}
\begin{barticle}
\bauthor{\bsnm{Shirahama}, \binits{K.}},
\bauthor{\bsnm{Kubota}, \binits{M.}},
\bauthor{\bsnm{Ogawa}, \binits{S.}},
\bauthor{\bsnm{Wada}, \binits{N.}},
\bauthor{\bsnm{Watanabe}, \binits{T.}}:
\batitle{Size-dependent {Kosterlitz}-{Thouless} superfluid transition in thin {$^4$He} films adsorbed on porous glasses}.
\bjtitle{Physical Review Letters}
\bvolume{64}(\bissue{13}),
\bfpage{1541}--\blpage{1544}
(\byear{1990})
\doiurl{10.1103/PhysRevLett.64.1541}
\end{barticle}
\endbibitem

\bibitem[\protect\citeauthoryear{Tilley and Tilley}{2003}]{tilley_2003}
\begin{bbook}
\bauthor{\bsnm{Tilley}, \binits{D.R.}},
\bauthor{\bsnm{Tilley}, \binits{J.}}:
\bbtitle{Superfluidity and Superconductivity},
\bedition{3. ed., repr} edn.
\bsertitle{Graduate student series in physics}.
\bpublisher{Inst. of Physics Publ},
\blocation{Bristol}
(\byear{2003})
\end{bbook}
\endbibitem

\bibitem[\protect\citeauthoryear{Perron et~al.}{2010}]{perron_2010}
\begin{barticle}
\bauthor{\bsnm{Perron}, \binits{J.K.}},
\bauthor{\bsnm{Kimball}, \binits{M.O.}},
\bauthor{\bsnm{Mooney}, \binits{K.P.}},
\bauthor{\bsnm{Gasparini}, \binits{F.M.}}:
\batitle{Coupling and proximity effects in the superfluid transition in {$^4$He} dots}.
\bjtitle{Nature Physics}
\bvolume{6}(\bissue{7}),
\bfpage{499}--\blpage{502}
(\byear{2010})
\doiurl{10.1038/nphys1671}
\end{barticle}
\endbibitem

\bibitem[\protect\citeauthoryear{Perron and Gasparini}{2012}]{perron_2012}
\begin{barticle}
\bauthor{\bsnm{Perron}, \binits{J.K.}},
\bauthor{\bsnm{Gasparini}, \binits{F.M.}}:
\batitle{Critical {Point} {Coupling} and {Proximity} {Effects} in {$^4$He} at the {Superfluid} {Transition}}.
\bjtitle{Physical Review Letters}
\bvolume{109}(\bissue{3}),
\bfpage{035302}
(\byear{2012})
\doiurl{10.1103/PhysRevLett.109.035302}
\end{barticle}
\endbibitem

\bibitem[\protect\citeauthoryear{Perron and Gasparini}{2013}]{perron_2013}
\begin{barticle}
\bauthor{\bsnm{Perron}, \binits{J.K.}},
\bauthor{\bsnm{Gasparini}, \binits{F.M.}}:
\batitle{Specific {Heat} and {Superfluid} {Density} of {$^4$He} near {$T_\lambda$} of a 33.6 nm {Film} {Formed} {Between} {Si} {Wafers}}.
\bjtitle{Journal of Low Temperature Physics}
\bvolume{171}(\bissue{5}),
\bfpage{589}--\blpage{598}
(\byear{2013})
\doiurl{10.1007/s10909-012-0795-0}
\end{barticle}
\endbibitem

\bibitem[\protect\citeauthoryear{Ambegaokar et~al.}{1978}]{ambegaokar_1978}
\begin{barticle}
\bauthor{\bsnm{Ambegaokar}, \binits{V.}},
\bauthor{\bsnm{Halperin}, \binits{B.I.}},
\bauthor{\bsnm{Nelson}, \binits{D.R.}},
\bauthor{\bsnm{Siggia}, \binits{E.D.}}:
\batitle{Dissipation in {Two}-{Dimensional} {Superfluids}}.
\bjtitle{Physical Review Letters}
\bvolume{40}(\bissue{12}),
\bfpage{783}--\blpage{786}
(\byear{1978})
\doiurl{10.1103/PhysRevLett.40.783}
\end{barticle}
\endbibitem

\bibitem[\protect\citeauthoryear{Ambegaokar et~al.}{1980}]{ambegaokar_1980}
\begin{barticle}
\bauthor{\bsnm{Ambegaokar}, \binits{V.}},
\bauthor{\bsnm{Halperin}, \binits{B.I.}},
\bauthor{\bsnm{Nelson}, \binits{D.R.}},
\bauthor{\bsnm{Siggia}, \binits{E.D.}}:
\batitle{Dynamics of superfluid films}.
\bjtitle{Physical Review B}
\bvolume{21}(\bissue{5}),
\bfpage{1806}--\blpage{1826}
(\byear{1980})
\doiurl{10.1103/PhysRevB.21.1806}
\end{barticle}
\endbibitem

\bibitem[\protect\citeauthoryear{Rojas and Davis}{2015}]{rojas_2015}
\begin{barticle}
\bauthor{\bsnm{Rojas}, \binits{X.}},
\bauthor{\bsnm{Davis}, \binits{J.P.}}:
\batitle{Superfluid nanomechanical resonator for quantum nanofluidics}.
\bjtitle{Physical Review B}
\bvolume{91}(\bissue{2}),
\bfpage{024503}
(\byear{2015})
\doiurl{10.1103/PhysRevB.91.024503}
\end{barticle}
\endbibitem

\bibitem[\protect\citeauthoryear{Souris et~al.}{2017}]{souris_2017}
\begin{barticle}
\bauthor{\bsnm{Souris}, \binits{F.}},
\bauthor{\bsnm{Rojas}, \binits{X.}},
\bauthor{\bsnm{Kim}, \binits{P.H.}},
\bauthor{\bsnm{Davis}, \binits{J.P.}}:
\batitle{Ultralow-{Dissipation} {Superfluid} {Micromechanical} {Resonator}}.
\bjtitle{Physical Review Applied}
\bvolume{7}(\bissue{4}),
\bfpage{044008}
(\byear{2017})
\doiurl{10.1103/PhysRevApplied.7.044008}
\end{barticle}
\endbibitem

\bibitem[\protect\citeauthoryear{Brooks and Donnelly}{1977}]{brooks_1977}
\begin{barticle}
\bauthor{\bsnm{Brooks}, \binits{J.S.}},
\bauthor{\bsnm{Donnelly}, \binits{R.J.}}:
\batitle{The calculated thermodynamic properties of superfluid helium-4}.
\bjtitle{Journal of Physical and Chemical Reference Data}
\bvolume{6}(\bissue{1}),
\bfpage{51}--\blpage{104}
(\byear{1977})
\doiurl{10.1063/1.555549}
\end{barticle}
\endbibitem

\bibitem[\protect\citeauthoryear{Padmore}{1974}]{padmore_1974}
\begin{barticle}
\bauthor{\bsnm{Padmore}, \binits{T.C.}}:
\batitle{Two-{Dimensional} {Rotons}}.
\bjtitle{Physical Review Letters}
\bvolume{32}(\bissue{15}),
\bfpage{826}--\blpage{829}
(\byear{1974})
\doiurl{10.1103/PhysRevLett.32.826}
\end{barticle}
\endbibitem

\bibitem[\protect\citeauthoryear{Kiewiet et~al.}{1975}]{Kiewiet1975}
\begin{barticle}
\bauthor{\bsnm{Kiewiet}, \binits{C.W.}},
\bauthor{\bsnm{Hall}, \binits{H.E.}},
\bauthor{\bsnm{Reppy}, \binits{J.D.}}:
\batitle{Superfluid density in porous vycor glass}.
\bjtitle{Physical Review Letters}
\bvolume{35}(\bissue{19}),
\bfpage{1286}--\blpage{1289}
(\byear{1975})
\doiurl{10.1103/PhysRevLett.35.1286}
\end{barticle}
\endbibitem

\bibitem[\protect\citeauthoryear{Donnelly and Barenghi}{1998}]{donnelly_1998}
\begin{barticle}
\bauthor{\bsnm{Donnelly}, \binits{R.J.}},
\bauthor{\bsnm{Barenghi}, \binits{C.F.}}:
\batitle{The {Observed} {Properties} of {Liquid} {Helium} at the {Saturated} {Vapor} {Pressure}}.
\bjtitle{Journal of Physical and Chemical Reference Data}
\bvolume{27}(\bissue{6}),
\bfpage{1217}--\blpage{1274}
(\byear{1998})
\doiurl{10.1063/1.556028}
\end{barticle}
\endbibitem

\bibitem[\protect\citeauthoryear{Singasaas and Ahlers}{1984}]{singasaas_1984}
\begin{barticle}
\bauthor{\bsnm{Singasaas}, \binits{A.}},
\bauthor{\bsnm{Ahlers}, \binits{G.}}:
\batitle{Universality of static properties near the superfluid transition in {$^4$He}}.
\bjtitle{Physical Review B}
\bvolume{30}(\bissue{9}),
\bfpage{5103}--\blpage{5115}
(\byear{1984})
\doiurl{10.1103/PhysRevB.30.5103}
\end{barticle}
\endbibitem

\bibitem[\protect\citeauthoryear{Chester and Eytel}{1976}]{chester_1976}
\begin{barticle}
\bauthor{\bsnm{Chester}, \binits{M.}},
\bauthor{\bsnm{Eytel}, \binits{L.}}:
\batitle{Some calculations regarding the characteristic length for superfluidity in liquid helium}.
\bjtitle{Physical Review B}
\bvolume{13}(\bissue{3}),
\bfpage{1069}--\blpage{1076}
(\byear{1976})
\doiurl{10.1103/PhysRevB.13.1069}
\end{barticle}
\endbibitem

\bibitem[\protect\citeauthoryear{Arrigoni et~al.}{2013}]{arrigoni_2013}
\begin{barticle}
\bauthor{\bsnm{Arrigoni}, \binits{F.}},
\bauthor{\bsnm{Vitali}, \binits{E.}},
\bauthor{\bsnm{Galli}, \binits{D.E.}},
\bauthor{\bsnm{Reatto}, \binits{L.}}:
\batitle{Excitation spectrum in two-dimensional superfluid {$^4$He}}.
\bjtitle{Low Temperature Physics}
\bvolume{39}(\bissue{9}),
\bfpage{793}--\blpage{800}
(\byear{2013})
\doiurl{10.1063/1.4821079}
\end{barticle}
\endbibitem

\bibitem[\protect\citeauthoryear{Agnolet et~al.}{1981}]{agnolet_thermal_1981}
\begin{barticle}
\bauthor{\bsnm{Agnolet}, \binits{G.}},
\bauthor{\bsnm{Teitel}, \binits{S.L.}},
\bauthor{\bsnm{Reppy}, \binits{J.D.}}:
\batitle{Thermal {Transport} in a $^4${He} {Film} at the {Kosterlitz}-{Thouless} {Transition}}.
\bjtitle{Physical Review Letters}
\bvolume{47}(\bissue{21}),
\bfpage{1537}--\blpage{1540}
(\byear{1981})
\doiurl{10.1103/PhysRevLett.47.1537}
\end{barticle}
\endbibitem

\bibitem[\protect\citeauthoryear{Novotný et~al.}{2026}]{novotny_2026}
\begin{barticle}
\bauthor{\bsnm{Novotný}, \binits{F.}},
\bauthor{\bsnm{Talíř}, \binits{M.}},
\bauthor{\bsnm{Varga}, \binits{E.}}:
\batitle{Decay of {Two}-{Dimensional} {Superfluid} {Turbulence} over {Pinning} {Surface}}.
\bjtitle{Physical Review Letters}
\bvolume{136}(\bissue{8}),
\bfpage{084003}
(\byear{2026})
\doiurl{10.1103/qrrj-p85y}
\end{barticle}
\endbibitem

\bibitem[\protect\citeauthoryear{Williams}{1987}]{Williams1987}
\begin{barticle}
\bauthor{\bsnm{Williams}, \binits{G.A.}}:
\batitle{Vortex-ring model of the superfluid $\lambda$ transition}.
\bjtitle{Phys. Rev. Lett.}
\bvolume{59}(\bissue{17}),
\bfpage{1926}--\blpage{1929}
(\byear{1987})
\doiurl{10.1103/PhysRevLett.59.1926}
\end{barticle}
\endbibitem

\bibitem[\protect\citeauthoryear{Gasparini and Mehta}{1998}]{gasparini_1998}
\begin{barticle}
\bauthor{\bsnm{Gasparini}, \binits{F.M.}},
\bauthor{\bsnm{Mehta}, \binits{S.}}:
\batitle{Temperature {Oscillations} of an {Adiabatic} {Superleak}}.
\bjtitle{Journal of Low Temperature Physics}
\bvolume{110}(\bissue{1}),
\bfpage{293}--\blpage{298}
(\byear{1998})
\doiurl{10.1023/A:1022528631749}
\end{barticle}
\endbibitem

\bibitem[\protect\citeauthoryear{Kimball and Gasparini}{2001}]{kimball_2001}
\begin{barticle}
\bauthor{\bsnm{Kimball}, \binits{M.O.}},
\bauthor{\bsnm{Gasparini}, \binits{F.M.}}:
\batitle{Superfluid {Fraction} of {$^{3}$He} - {$^4$He} {Mixtures} {Confined} at 0.0483 {$\mathrm{\mu}$m} between {Silicon} {Wafers}}.
\bjtitle{Physical Review Letters}
\bvolume{86}(\bissue{8}),
\bfpage{1558}--\blpage{1561}
(\byear{2001})
\doiurl{10.1103/PhysRevLett.86.1558}
\end{barticle}
\endbibitem

\bibitem[\protect\citeauthoryear{Yu et~al.}{1989}]{yu_1989}
\begin{barticle}
\bauthor{\bsnm{Yu}, \binits{Y.Y.}},
\bauthor{\bsnm{Finotello}, \binits{D.}},
\bauthor{\bsnm{Gasparini}, \binits{F.M.}}:
\batitle{Finite-size scaling and the convective conductance and specific heat of planar helium films near the superfluid transition}.
\bjtitle{Physical Review B}
\bvolume{39}(\bissue{10}),
\bfpage{6519}--\blpage{6526}
(\byear{1989})
\doiurl{10.1103/PhysRevB.39.6519}
\end{barticle}
\endbibitem

\bibitem[\protect\citeauthoryear{Ganshin et~al.}{2006}]{ganshin_2006}
\begin{barticle}
\bauthor{\bsnm{Ganshin}, \binits{A.}},
\bauthor{\bsnm{Scheidemantel}, \binits{S.}},
\bauthor{\bsnm{Garcia}, \binits{R.}},
\bauthor{\bsnm{Chan}, \binits{M.H.W.}}:
\batitle{Critical {Casimir} {Force} in {$^4$He} {Films}: {Confirmation} of {Finite}-{Size} {Scaling}}.
\bjtitle{Physical Review Letters}
\bvolume{97}(\bissue{7}),
\bfpage{075301}
(\byear{2006})
\doiurl{10.1103/PhysRevLett.97.075301}
\end{barticle}
\endbibitem

\bibitem[\protect\citeauthoryear{Novotn\'{y} et~al.}{2026}]{zenodo}
\begin{botherref}
\oauthor{\bsnm{Novotn\'{y}}, \binits{F.}},
\oauthor{\bsnm{Tal\'{i}\v{r}}, \binits{M.}},
\oauthor{\bsnm{Szalai}, \binits{B.}},
\oauthor{\bsnm{Varga}, \binits{E.}}:
Dataset for the manuscript, containing raw data and analysis scripts that produce figures in the paper.
(2026)
\doiurl{10.5281/zenodo.18836903}
\end{botherref}
\endbibitem

\bibitem[\protect\citeauthoryear{Doolin}{2019}]{doolin2019}
\begin{botherref}
\oauthor{\bsnm{Doolin}, \binits{C.}}:
Integrated optical and mechanical resonators for evanescent field sensing.
PhD thesis,
University of Alberta
(2019)
\end{botherref}
\endbibitem

\bibitem[\protect\citeauthoryear{Shook et~al.}{2020}]{shook_stabilized_2020}
\begin{barticle}
\bauthor{\bsnm{Shook}, \binits{A.J.}},
\bauthor{\bsnm{Vadakkumbatt}, \binits{V.}},
\bauthor{\bsnm{Senarath~Yapa}, \binits{P.}},
\bauthor{\bsnm{Doolin}, \binits{C.}},
\bauthor{\bsnm{Boyack}, \binits{R.}},
\bauthor{\bsnm{Kim}, \binits{P.H.}},
\bauthor{\bsnm{Popowich}, \binits{G.G.}},
\bauthor{\bsnm{Souris}, \binits{F.}},
\bauthor{\bsnm{Christani}, \binits{H.}},
\bauthor{\bsnm{Maciejko}, \binits{J.}},
\bauthor{\bsnm{Davis}, \binits{J.P.}}:
\batitle{Stabilized {Pair} {Density} {Wave} via {Nanoscale} {Confinement} of {Superfluid} $^3${He}}.
\bjtitle{Physical Review Letters}
\bvolume{124}(\bissue{1}),
\bfpage{015301}
(\byear{2020})
\doiurl{10.1103/PhysRevLett.124.015301}
\end{barticle}
\endbibitem

\bibitem[\protect\citeauthoryear{Varga and Davis}{2021}]{varga_electromechanical_2021}
\begin{barticle}
\bauthor{\bsnm{Varga}, \binits{E.}},
\bauthor{\bsnm{Davis}, \binits{J.P.}}:
\batitle{Electromechanical feedback control of nanoscale superflow}.
\bjtitle{New Journal of Physics}
\bvolume{23}(\bissue{11}),
\bfpage{113041}
(\byear{2021})
\doiurl{10.1088/1367-2630/ac37c6}
\end{barticle}
\endbibitem

\bibitem[\protect\citeauthoryear{Novotný et~al.}{2025}]{novotny_2025}
\begin{barticle}
\bauthor{\bsnm{Novotný}, \binits{F.}},
\bauthor{\bsnm{Talíř}, \binits{M.}},
\bauthor{\bsnm{Midlik}, \binits{v.}},
\bauthor{\bsnm{Varga}, \binits{E.}}:
\batitle{Critical behavior and multistability in quasi-two-dimensional turbulence}.
\bjtitle{Physical Review Fluids}
\bvolume{10}(\bissue{5}),
\bfpage{054605}
(\byear{2025})
\doiurl{10.1103/PhysRevFluids.10.054605}
\end{barticle}
\endbibitem

\bibitem[\protect\citeauthoryear{Varga et~al.}{2020}]{varga_2020}
\begin{barticle}
\bauthor{\bsnm{Varga}, \binits{E.}},
\bauthor{\bsnm{Vadakkumbatt}, \binits{V.}},
\bauthor{\bsnm{Shook}, \binits{A.J.}},
\bauthor{\bsnm{Kim}, \binits{P.H.}},
\bauthor{\bsnm{Davis}, \binits{J.P.}}:
\batitle{Observation of {Bistable} {Turbulence} in {Quasi}-{Two}-{Dimensional} {Superflow}}.
\bjtitle{Physical Review Letters}
\bvolume{125}(\bissue{2}),
\bfpage{025301}
(\byear{2020})
\doiurl{10.1103/PhysRevLett.125.025301}
\end{barticle}
\endbibitem

\bibitem[\protect\citeauthoryear{López-Núñez et~al.}{2025}]{lopez_2025}
\begin{barticle}
\bauthor{\bsnm{López-Núñez}, \binits{D.}},
\bauthor{\bsnm{Torras-Coloma}, \binits{A.}},
\bauthor{\bsnm{Portell-Montserrat}, \binits{Q.}},
\bauthor{\bsnm{Bertoldo}, \binits{E.}},
\bauthor{\bsnm{Cozzolino}, \binits{L.}},
\bauthor{\bsnm{Ummarino}, \binits{G.A.}},
\bauthor{\bsnm{Zaccone}, \binits{A.}},
\bauthor{\bsnm{Rius}, \binits{G.}},
\bauthor{\bsnm{Martínez}, \binits{M.}},
\bauthor{\bsnm{Forn-Díaz}, \binits{P.}}:
\batitle{Superconducting penetration depth of aluminum thin films}.
\bjtitle{Superconductor Science and Technology}
\bvolume{38}(\bissue{9}),
\bfpage{095004}
(\byear{2025})
\doiurl{10.1088/1361-6668/adf360}
\end{barticle}
\endbibitem

\bibitem[\protect\citeauthoryear{Clements et~al.}{1996}]{clements_1996}
\begin{barticle}
\bauthor{\bsnm{Clements}, \binits{B.E.}},
\bauthor{\bsnm{Godfrin}, \binits{H.}},
\bauthor{\bsnm{Krotscheck}, \binits{E.}},
\bauthor{\bsnm{Lauter}, \binits{H.J.}},
\bauthor{\bsnm{Leiderer}, \binits{P.}},
\bauthor{\bsnm{Passiouk}, \binits{V.}},
\bauthor{\bsnm{Tymczak}, \binits{C.J.}}:
\batitle{Excitations in a thin liquid {$^4$He} film from inelastic neutron scattering}.
\bjtitle{Physical Review B}
\bvolume{53}(\bissue{18}),
\bfpage{12242}--\blpage{12252}
(\byear{1996})
\doiurl{10.1103/PhysRevB.53.12242}
\end{barticle}
\endbibitem

\bibitem[\protect\citeauthoryear{Mitin and Kholevchuk}{2025}]{mitin_2025}
\begin{barticle}
\bauthor{\bsnm{Mitin}, \binits{V.F.}},
\bauthor{\bsnm{Kholevchuk}, \binits{V.V.}}:
\batitle{Miniature resistance thermometers based on {Ge}/{GaAs} films for use in the 1.5 to 300 {K}, 70 to 400 {K}, and 200 to 500 {K} temperature ranges}.
\bjtitle{Low Temperature Physics}
\bvolume{51}(\bissue{10}),
\bfpage{1293}--\blpage{1299}
(\byear{2025})
\doiurl{10.1063/10.0039431}
\end{barticle}
\endbibitem

\end{thebibliography}


\appendix

\section{Measurement scheme and signal fitting}

\begin{figure}[ht!]
    \centering
    \includegraphics{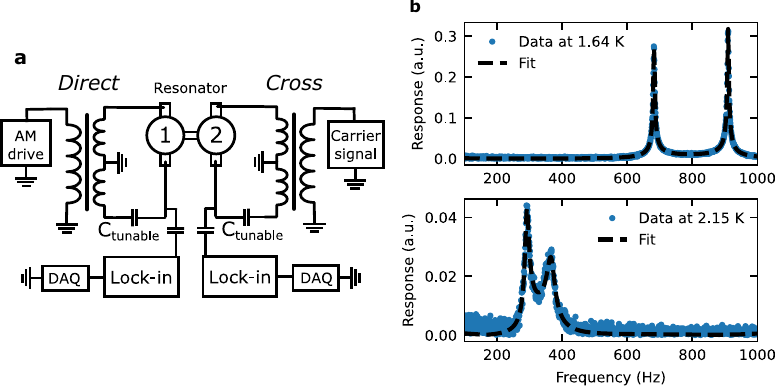}
    \caption{(a) Scheme of the actual measurement circuit. (b) Example of the measured signal with a double lorentzian fit for two different temperatures. The measurement of the resonator with the 10~nm nanochannel.}
    \label{fig:circuit}
\end{figure}

In Fig.\ref{fig:circuit}(a) we show the measurement scheme used in the experiment. A pair of aluminium electrodes was evaporated in each of the resonator basins, forming a capacitor with the liquid \4He acting as the dielectric. Each basin is connected to a separate measurement bridge, which were tuned prior to the measurements. The pressure fluctuation in the basin on resonance results in capacitance fluctuation and thus a nonzero bedige current detected by a lock-in amplifier. 

The resonant motion was directly driven in only one basin (\emph{direct}), while in the second basin the motion was induced by flow through the connecting nanochannel. Since the temperature during the measurements was not actively stabilized but instead changed very slowly with time (either increasing or decreasing), we employed a pulse method to excite the resonance \cite{doolin2019,shook_stabilized_2020}. In this method, an approximately 3~s-long voltage pulse $v(t)$ sweeping over a given frequency range was applied to the direct basin. The time-domain response was recorded from both basins simultaneously using a fast data acquisition (DAQ) card. The resonant response was determined as the ratio of the Fourier transforms of the response and the pulse, which allowed us to rapidly record the temperature dependence of the acoustic resonance. Since the frequencies of interest are relatively low, to suppress external line noise that overwhelms the studied response, we employed a double-demodulation technique as in ref.~\cite{varga_electromechanical_2021}: the acoustic resonance is observed as amplitude modulation sidebands on a relatively high frequency carrier (approximately 70 kHz, chosen arbitrarily in a region away from spurious noise peaks). These carrier tones were applied to both basins separately and then demodulated using a Stanford Research SR830 lock-in amplifiers before being captured by the DAQ.

The recorded response consists of four signals: in-phase ($X$) and quadrature ($Y$) with the carrier tone on the direct and cross basin. After Fourier transform, these result in four complex spectra that are separately fit using a complex double-Lorentzian function
\begin{equation}
    \tilde{s}(f) = \dfrac{f_{0}A_0 \gamma_0}{f_0^2 - f^2 + if\gamma_0}e^{i \phi_0} + \dfrac{f_{1}A_1 \gamma_1}{f_1^2 - f^2 + if\gamma_1}e^{i \phi_1} + (\textrm{background}),
\end{equation}
The background was modelled by a complex fourth-order polynomial in $f$. The parameters $f_0$ and $f_1$ denote the resonance frequencies, $A_0$ and $A_1$ the amplitudes, $\gamma_0$ and $\gamma_1$ widths of the resonance peaks, and $\phi_0$ and $\phi_1$ the phases, all of which are adjustable parameters.

An example of the measured signal together with the corresponding fit is shown in Fig.\ref{fig:circuit}(b), showing the absolute value of the spectrum obtained from $X_\mathrm{cross}$. With increasing temperature, a gradual merging of the resonance modes is observed. After the modes merge, the cross signal disappears. For the direct signal, however, we continue to fit the single remaining peak using a single-Lorentzian function, which provides sufficient resolution up to $\Tl$.

\section{Extrapolation of $f_0(0)$ and $f_1(0)$ and roton excitations.}

\begin{figure}[ht!]
    \centering
    \includegraphics{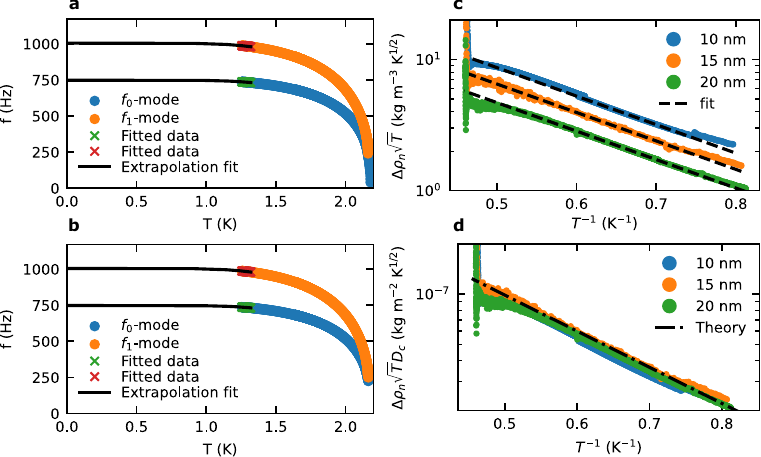}
    \caption{(a,b) Obtained resonance frequency by the double-lorentzian fit to measured spectra against temperature for both type of signals. The full black lines mark the extrapolation fit to the low temperature part of the data. The shown example is for the resonator with the 10~nm nanochannel. (c) Normal fluid enhancement $\Delta \rho_n$ in the nanochannel rescaled by $\sqrt{T}$ against $1/T$ for all three confinements. The black dashed lines mark the linear fit giving the roton gap $\Delta/k_B \approx5$~K after slight correction of fitted $f_0(0)$ and $f_1(0)$. (d) Data from the graph (c) rescaled by the confinement $D_c$ compared to the theoretical prediction for density of 2D rotons the Eq.(4) in the main text for $k_0 \approx 1.8 \textrm{\AA}^{-1} $, $m^{*} = 0.2m_{4He}$ and $\Delta/k_B = 5$~K \cite{padmore_1974,chester_1976,arrigoni_2013}.}
    \label{fig:roton}
\end{figure}

The zero-temperature resonance frequencies $f_0(0)$ and $f_1(0)$ are needed for the calculation of $\rhosb$ and $\rhosc$, which have to be determined by extrapolation below the the lowest temperature reachable by pumping on the saturated vapour in the present experiment, slightly above 1.2~K. At these temperatures, the superfluid density of \4He is still varying, leading to changes in the resonance frequencies. The low-temperature part of the data, see Fig.\ref{fig:roton}(a), are fit to (see main text and \cite{varga_2022})
\begin{align}
    f_0 &= \dfrac{1}{2\pi} \sqrt{\alpha\dfrac{2wHk_p}{l(2A^2 + \beta A H \chi k_p)} \dfrac{\rho_{sb}}{\rho^2}}, \label{f0} \\
    f_1 &= \dfrac{1}{2\pi} \sqrt{\alpha\dfrac{2wHk_p}{l(2A^2 + \beta A H \chi k_p)}}  \sqrt{\dfrac{\rho_{sb}}{\rho^2} + \dfrac{w_c}{w} \dfrac{D_c}{H} \dfrac{l}{l_c}\dfrac{\rho_{sc}}{\rho^2}}. \label{f1}
\end{align}
Here, $\alpha$ and $\beta$ are the fitted parameters similar to refs.~\cite{souris_2017, novotny_2025}. The remaining parameters are: $w$ and $w_c$, the widths of the filling channels and the nanochannel, $l$ and $l_c$ their corresponding lengths, $H$ and $D_c$ the heights of the basins and the nanochannel, and $A$ is the area of the basin. The parameter $k_p$ is the stiffness of the basin, for which we use a typical value of $10^7$~N/m \cite{souris_2017, varga_2020, novotny_2025}, and $\chi$ is the isothermal compressibility of \4He, which is taken from \cite{brooks_1977}. The temperature dependence enters into the Eqs.~\eqref{f0}, \eqref{f1} through $\rhosb$, $\rhosc$ and $\chi$. For simplicity, we assume $\rhosb \approx \rhosc$ at low temperatures. Although this approximation is not strictly correct, it has a negligible effect on the low-temperature fits.

To support the hypothesis that 2D rotons are responsible for the enhancement of the normal-fluid component in the nanochannel, defined as $\Delta\rho_n = \rhosb - \rhosc$, we plot the logarithm of the temperature-rescaled enhancement $\Delta\rho_n \sqrt{T}$ s a function of inverse temperature $1/T$, as shown in Fig.\ref{fig:roton}(b). According to the Eq.(4) in the main text, the negative slope of this dependence corresponds to the roton gap $\Delta/k_B$. To ensure consistency with a previous study of double-basin Helmholtz resonators \cite{varga_2022}, in which the extracted gap was close to the expected value of 5~K \cite{padmore_1974}, we applied a slight correction to the extrapolated zero-temperature frequencies $f_0(0)$ and $f_1(0)$. The correction was on the order of a units Hz, corresponding to less than one percent. This adjustment was necessary because we observed a small jump in the measured resonance frequencies at $1/T \approx 0.75$~K$^{-1}$ ($T \approx 1.33$~K). We attribute this feature to the superconducting transition of the thin aluminum layer, whose critical temperature can be close to 1.3~K \cite{lopez_2025}. After the transition, the resistive aluminum causes additional Joule heating, which slightly modifies the superfluid density and, consequently, the resonance frequencies.

The additional jump in the 10~nm data appears due to stitching together data sets acquired with slightly different settings of the high-frequency carrier signal, primarily differing in amplitude and with different liquid helium heights in the cryostat, resulting in different hydrostatic pressure. Such changes affect the effective basin stiffness $k_p$ and total fluid density, leading to small shifts in the resonance frequencies. Thus, the fits shown in Fig.\ref{fig:roton}(b) were restricted to the range $0.6~\mathrm{K}^{-1} < 1/T <  0.7~\mathrm{K}^{-1}$, with the roton gap treated as a fitting parameter and yielding a value of 5~K.

By further rescaling the $\Delta\rho_n$ by the confinement $D_c$, converting the three-dimensional roton density to a two-dimensional density averaged across the channel height, the data collapse onto a single line, as shown in Fig.\ref{fig:roton}(c). We directly compare the collapsed data with the theoretical expression using parameters $k_0 \approx 1.8 \textrm{\AA}^{-1}$, $m^{*} = 0.2m$ and $\Delta/k_B$ obtained from numerical simulations \cite{padmore_1974,chester_1976,arrigoni_2013} and neutron scattering \cite{clements_1996}. We find excellent agreement between our measurements and the theoretical prediction.

Finally, we emphasize that the small corrections to $f_0(0)$ and $f_1(0)$ did not affect the high-temperature behaviour of $\rhosb$ and $\rhosc$ near $\Tl$.

\section{Absolute temperature correction: $\Tl$ fitting}

\begin{figure}[ht!]
    \centering
    \includegraphics{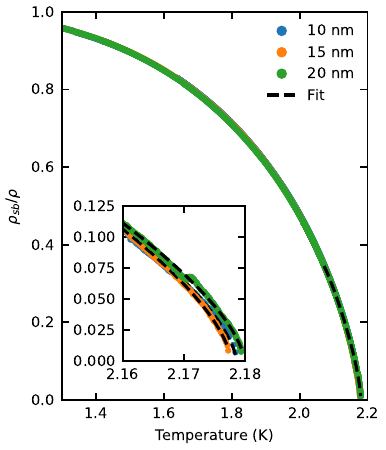}
    \caption{Ratio of the bulk superfluid density and total density obtained from the $f_0$-mode of the direct-signal against temperature measured by a resistive thermometer \cite{mitin_2025}. The inset show the high temperature behaviour with the fits using the formula \eqref{ratio}.}
    \label{fig:Tl}
\end{figure}

The accurate determination of $\Tl$ is crucial for analysing the high-temperature power-law behaviour of $\rhosb$, where a straight line with slope 0.67 is expected in the log–log plots shown in insets of Fig.~2(a,b) in the main article and for identifying the correct temperatures of the KT transition region in the case of $\rhosc$. For the temperature measurement we used a semiconducting GeAs thermometer (TTRG) \cite{mitin_2025}, which was calibrated before the measurement using the tabulated saturated vapour pressure \cite{donnelly_1998}. However, the resistance of the thermometer may vary between separate measurements at the same temperature. According to the manufacturer, the stability is no better than 5~mK \cite{mitin_2025}. Therefore, we determined $\Tl$ separately for each resonator by fitting the empirical formula \cite{singasaas_1984}
\begin{equation}\label{ratio}
    \dfrac{\rhosb}{\rho} = k_0 ( 1 + k_1 |t|)|t|^{\xi}(1+k_2|t|^{\Delta}),
\end{equation}
to the measured data. Here $t = 1 - T/T_{\lambda}$, $\xi = 0.6717$, $\Delta = 0.5$ and $k_0$, $k_1$, $k_2$ and $\Tl$ are fitted parameters. This formula describes the ratio $\rhosb/\rho$ as a function of the reduced temperature $t$ with sufficient accuracy for $T \geq 2.072$~K. The ratios obtained from the measured frequencies of the $f_0$-mode of the direct signal are shown in Fig.\ref{fig:Tl} together with the corresponding fits. In the case of the 20~nm resonator, a small bump is observed at $\approx$2.17~K which is caused by a change in $f_0$ due to modification of the resonant-mode geometry following blockage of the nanochannel. This feature can slightly perturb the fitted $\Tl$, but the final high temperature behaviour of $\rhosb$ is still in good agreement with the theory, as one can see in insets of Fig.~2(a,b) in the main text, where fitted $\Tl$ values were used.

\section{Background dissipation subtraction}

To correctly fit the dynamical AHNS theory to the dissipation peak measured by the $f_1$ mode using the cross signal (resonance frequency $f_{1c}$), it is necessary to subtract the background dissipation originating from flow in other parts of the chip. This background dissipation arises from viscous friction between the substrate and the normal component of \4He, as well as from thermal losses through the substrate and the filling channels \cite{souris_2017}.

\begin{figure}
    \centering
    \includegraphics{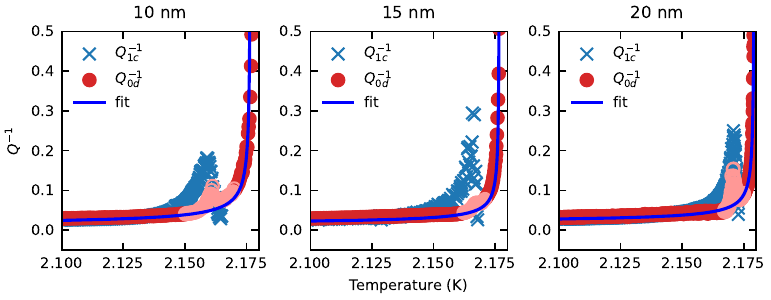}
    \caption{Dependence of the inverse Q-factor on the temperature close to the $\lambda$-point. Crosses mark $Q^{-1}$ of the antisymmetric mode, while circles the fundamental mode. The blue line is the power law fit to the dark red part of the data. Subscripts c and d refer to cross and direct signals, respectively.}
    \label{fig:invQ}
\end{figure}

In our device, the symmetry of the resonator ensures that all resonance modes experience the same background dissipation. Therefore, the dissipation associated with the KT transition can be isolated by subtracting the dissipation given by peak widths of one mode from another. Or, in terms of $Q^{-1}$, after appropriate rescaling by the mode frequencies. Specifically, we subtract from the inverse quality factor of the first mode $Q^{-1}_{1c}$, the rescaled inverse quality factor of the fundamental mode, $Q^{-1}_{0d}f_{0d}/f_{1c}$, where $f_{0d}$ is the resonance frequency of the $f_0$ mode, see the Eq.(5) and Fig.3(a) in the main text (subscripts c and d refer to cross and direct signal, respectively). We chose $f_0$ mode, because it could be tracked up to $\Tl$. This mode is also partially sensitive to the KT transition, which manifests in smaller dissipation peak, which was cut and the missing range of $Q^{-1}_{0d}$ was interpolated using a power law fit of the form $A|T - T_{\lambda}|^{-0.67} + B$. This procedure is demonstrated in Fig.\ref{fig:invQ}.

\section{AFM scans}
\begin{figure}[h!]
    \centering
    \includegraphics{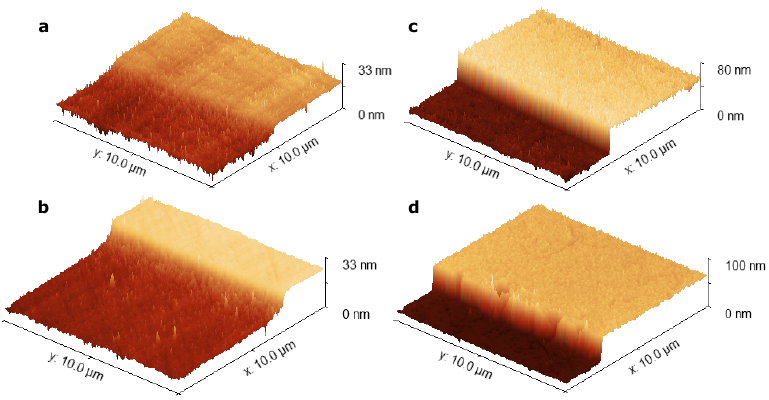}
    \caption{(a,b) AFM scans of the edge of half of the 5~nm and 10~nm high nanochannel. (c,d) AFM scan of the edge of the aluminium electrode.}
    \label{fig:AFM_scan}
\end{figure}

The resonators are composed of two identical half chips which enclose the nanofluidic cavity. To characterize the roughness of the surface in contact with helium we performed the atomic force microcopy scans on the half chip with nominally 5~nm and 10~nm high half-nanochannel. Since the measurement is destructive, the chips used for the AFM scan were not used in the experiments, but are from the same wafers. We scanned the edges of the nanochannel (Fig.\ref{fig:AFM_scan}(a,b)) and the edges of the aluminium electrode (Fig.\ref{fig:AFM_scan}(c,d)). 

The scanned heights are between 4~nm to 5~nm (5~nm half-nanochannel) and between 11~nm to 12~nm (10~nm half-nanochannel). The root mean square roughness of the 10~nm half-nanochannel surface measured 2~$\mu m$ from the edge is $R_{RMS}\approx1.0$~nm and 1.2~nm for the 5~nm half-nanochannel. The RMS roughness of the aluminium is then for both resonators similar $R_{RMS} \approx 2.3$~nm, when we avoid larger random defects in the aluminium layer.

\end{document}